\documentclass[reprint,longbibliography,preprintnumbers,nofootinbib,amsmath,amssymb,aps,twocolumns,prd]{revtex4-2}
\usepackage{graphicx}
\usepackage{dcolumn}
\usepackage{bm}
\usepackage{color}
\usepackage{amssymb}
\usepackage{pifont}
\definecolor{green2}{rgb}{0.0, 0.5, 0.0}
\usepackage[dvipsnames]{xcolor}
\usepackage[colorlinks = true,
            linkcolor = blue,
            urlcolor  = blue,
            citecolor = green,
            anchorcolor = blue]{hyperref}
\usepackage{verbatim}
\makeatletter
\usepackage{multirow, graphicx,amssymb,url,mathrsfs,amsmath}
\usepackage{amsxtra,amstext,latexsym,dsfont,amsfonts}
\usepackage{color,eucal}
\usepackage[dvipsnames]{xcolor}
\usepackage{float}
\usepackage{colortbl}
\usepackage{multirow}
\usepackage{tikz}
\usetikzlibrary{tikzmark}
\usetikzlibrary{arrows}

\newcommand{\Rmnum}[1]{\expandafter\@slowromancap\romannumeral #1@}
\usepackage{tabularx, array}
\newcolumntype{L}[1]{>{\raggedright\arraybackslash}p{#1}}
\newcolumntype{C}[1]{>{\centering\arraybackslash}p{#1}}
\newcolumntype{R}[1]{>{\raggedleft\arraybackslash}p{#1}}
\makeatother

\makeatletter
\newcommand{\myBig}{\bBigg@{1.75}}
\makeatother

\begin{document}
\title{\Large Phase transitions in a holographic s+d model from the 4D Einstein-Gauss-Bonnet gravity}
\author{Ting-Rui Liao}
\email{These authors contributed equally to this work and should be considered co-first authors.}
\affiliation{Center for Gravitation and Astrophysics, Kunming University of Science and Technology, Kunming 650500, China}
\author{Xin Zhao}
\email{These authors contributed equally to this work and should be considered co-first authors.}
\affiliation{Center for Gravitation and Astrophysics, Kunming University of Science and Technology, Kunming 650500, China}
\author{Hui Zeng}
\email{zenghui@kust.edu.cn}
\email{Authors to whom any correspondence should be addressed.}
\affiliation{Center for Gravitation and Astrophysics, Kunming University of Science and Technology, Kunming 650500, China}
\author{Zhang-Yu Nie}
\email{niezy@kust.edu.cn}
\email{Authors to whom any correspondence should be addressed.}
\affiliation{Center for Gravitation and Astrophysics, Kunming University of Science and Technology, Kunming 650500, China}
\date{\today}

\begin{abstract}
In this work, the phase structure of a holographic s+d model with quartic potential terms from the 4D Einstein-Gauss-Bonnet gravity is studied in the probe limit. We first show the $q_d-\mu$ phase diagram with a very small value of the Gauss-Bonnet coefficient $\alpha=1\times10^{-7}$ and in absence of the quartic terms to locate the suitable choice of the value of $q_d$, where the system admits coexistent s+d solutions. Then we consider various values of the Gauss-Bonnet coefficient $\alpha$ and present the $\alpha-\mu$ phase diagram to show the influence of the Gauss-Bonnet term on the phase structure. We also give an example of the reentrant phase transition which is also realized in the holographic s+s and s+p models. After that we confirm the universality of the influence of the quartic term with coefficient $\lambda_d$ on the d-wave solutions, which is similar to the case of s-wave and p-wave solutions previously studied in the s+p model. Finally we give the dependence of the special values of the quartic term coefficient $\lambda_d$ on the Gauss-Bonnet coefficient $\alpha$, below which the d-wave condensate grows to an opposite direction at the (quasi-)critical point, which is useful in realizing 1st order phase transitions in further studies of the holographic d-wave superfluids.
\end{abstract}
\maketitle
\tableofcontents
\section{Introduction}\label{Introduction}
In recent years, the anti-de Sitter/conformal field theory (AdS/CFT) duality~\cite{Maldacena:1997re} attracted a lot attention. It is regarded as an effective method for studying strongly coupled systems in QCD~\cite{Sakai:2004cn,Karch:2006pv} and condensed matter physics. One of the most successful applications of this duality is the holographic realization for superfluid phase transitions~\cite{Gubser:2008px,Hartnoll:2008vx}.
The holographic study on superfluid phase transitions is later extended from the s-wave order parameter to p-wave and d-wave orders~\cite{Gubser:2008wv,Cai:2013pda,Cai:2013aca,Chen:2010mk,Benini:2010pr} by introducing charged vector and tensor fields in the bulk. Based on these progresses, the competition and coexistence between these various orders were also investigated in the holographic models~\cite{Basu:2010fa,Cai:2013wma,Nie:2013sda,Musso:2013ija,Nishida:2014lta,Li:2014wca,Nie:2014qma,Nie:2015zia,Amado:2013lia,Arias:2016nww,Nie:2016pjt,Li:2017wbi,Nie:2020lop,Xia:2021pap,Zhang:2021vwp,Yang:2021mit}.

The interplay between the s-wave and d-wave orders are important in real superconductors such as cuprate and garnered considerable attention from condensed matter physicists. ~\cite{PhysRevB.37.3664,1995Muller,BealMonod1996PhysRevB,1997LiuPhysRevB,AngsulaGhosh1998,1999IoffeNatur398679I,2000Tsuei,2002Muller,2004Bussmann,2023Zhang}. In Ref.~\cite{PhysRevB.37.3664}, the authors show that the resonating-valence-bond mechanism can lead to s-wave, d-wave and s+d superconducting order parameters. It is claimed in Ref.~\cite{1995Muller} that the order parameter in the copper oxide superconductors should be coexistence of s-wave and d-wave condensates. Ref.~\cite{BealMonod1996PhysRevB} delves into the study of anisotropic s+d orders in both 2D and 3D contexts. In Ref.~\cite{1997LiuPhysRevB}, the authors explores the phase transitions between the s+d phase and the single condensate phases in the Van Hove scenario including the effects of orthorhombic distortion. Meanwhile, Ref.~\cite{AngsulaGhosh1998} realizes the phase transitions between the single condensate phase and the s+d phase in a BCS framework. Additionally, the S-D-S wave SQUID and Josephson junctions are investigated in Ref.~\cite{1999IoffeNatur398679I}. For a comprehensive review encompassing both theoretical and experimental studies on the pairing symmetry in cuprate superconductors, refer to Ref.~\cite{2000Tsuei}. For further advancements in this field, consult Refs.~\cite{2002Muller,2004Bussmann,2023Zhang}.

In recent years, the holographic model with both the s-wave and d-wave orders is also studied to explore possible universality~\cite{Nishida:2014lta,Li:2014wca}, where coexistence between the two orders are observed. It is found in Ref.~\cite{Nishida:2014lta} that the fourth power interaction term between the two charged fields control the width of the s+d coexisting phase. In a recent study on the holographic s+p model~\cite{Zhang:2021vwp}, it is found that the fourth power nonlinear terms have nontrivial and universal influence on the phase structure involving both the two orders. Therefore it is interesting to include the similar fourth power terms in the holographic s+d model to test this universality. 

From the gravity side, it is nature to consider higher curvature corrections to the standard Einstein gravity. The Gauss-Bonnet term is the best choice of the curvature square term, which do not bring in higher order derivative to the equations of motion~\cite{Lanczos:1938sf,Lovelock:1971yv}. According to the AdS/CFT correspondence, introducing the Gauss-Bonnet term corresponds to bring in corrections of large N expansion of CFTs in the strong coupling limit. In the study of holography, the Gauss-Bonnet term usually have nontrivial contribution to important results such as the universal KSS bound of the shear viscosity over entropy ratio~\cite{Kovtun:2004de,Brigante:2007nu,Brigante:2008gz,Cai:2008ph}, as well as the frequency gap in the holographic superconductor models~~\cite{Gregory:2009fj,Pan:2009xa,Barclay:2010up,Brihaye:2010mr,Gregory:2010yr, Cai:2010cv,Kanno:2011cs,Pan:2011ah,Li:2011xja,Lu:2016smd,Liu:2016utt}. However, the Gauss-Bonnet term is a topological invariant in 4 dimensions, and is believed to be only nontrivial in the gravity bulk with 5 or more spacetime dimensions. Until a recent study~\cite{Glavan:2019inb},
Glavan and Lin redefined the coupling constant and proposed a new Einstein-Gauss-Bonnet theory in four-dimensional spacetime, which is also generalized to Einstein-Lovelock gravity in 4D~\cite{Konoplya:2020qqh,Konoplya:2020juj,Casalino:2020kbt}. Consequently, we expect similar large N corrections from the Gauss-Bonnet term in the AdS4/CFT3 duality~\cite{Qiao:2020hkx}, which also introduce nontrivial influences on the holographic superconductors even in the probe limit.\cite{Qiao:2020hkx,Ghorai:2021uby,Pan:2021jii} 

In this study, we setup the holographic s+d superconductor model with 4th power nonlinear terms in the probe limit from the new 4D Einstein-Gauss-Bonnet gravity~\cite{Glavan:2019inb}, and study the various phase transitions. We will first study the impact of the Gauss-Bonnet term on the competition and coexistence of two order parameters. We also study the influence of the three different nonlinear terms on the phase transition to see whether the power of these terms on tuning the phase transitions are universal.

The rest of this article is structured as follows. In Sec.\ref{sect:2}, we provide the framework of the holographic model and the details of calculations as well as numerical works. In Sec.\ref{sect:3}, we give our main results from numerical calculations to show the influence of the Gauss-Bonnet term as well as the 4th power nonlinear terms. Finally, In Sec.\ref{sect:4}, we summarize our findings and outline potential avenues for future research.
\section{The holographic setup of the s+d mode from the 4D Einstein-Gauss-Bonnet gravity}  \label{sect:2}
The action of the holographic s+d superconductor model in the 4D Einstein-Gauss-Bonnet gravity is
\begin{align}
S=&S_M+S_G~, \\
S_{G}=& \frac{1}{2 \kappa_{g}^{2}} \int \text{d}^{4} x \sqrt{-g}\big(R-2 \Lambda+\bar{\alpha} R_{\mathrm{GB}}^{2}\big)~, \\
S_M= & \int \text{d}^{4} x \sqrt{-g}\big[-\frac{1}{4} F^{\mu \nu} F_{\mu \nu}+\mathcal{L}_{\mathrm{s}}+\mathcal{L}_{\mathrm{d}}\nonumber\\
&-\lambda_{sd}|\psi|^{2}(\left|\Phi_{\mu \nu}\right|^{2}-|\Phi|^{2})
\big]~,
\end{align}
where $S_G$ is the action of the gravity part and $S_M$ is the action of the matter part. ${R}_{\mathrm{GB}}^{2}=R^{2}-4 R_{\mu \nu} R^{\mu \nu}+R_{\mu \nu \rho \sigma} R^{\mu \nu \rho \sigma}$ is the Gauss-Bonnet term with the coefficient $\bar{\alpha}$. $S_M$ is a combination of a complex scalar and a complex tensor charged under the same U(1) gauge field with the field strength $F_{\mu \nu}=\nabla_{\mu} A_{\nu}-\nabla_{\nu} A_{\mu}$. The Laglagien of the s-wave part $\mathcal{L}_{\mathrm{s}}$ and the d-wave part $\mathcal{L}_{\mathrm{d}}$ are
\begin{align}
\mathcal{L}_{\mathrm{s}}= & -\left|D_{\mu} \Psi\right|^{2}-m_{\mathrm{s}}^{2}|\Psi|^{2}-\lambda_{s}|\Psi|^{4},\\
\mathcal{L}_{\mathrm{d}}= & -\left|D_{\rho} \Phi_{\mu \nu}\right|^{2}+2\left|D_{\mu} \Phi^{\mu \nu}\right|^{2}+\left|D_{\mu} \Phi\right|^{2}\nonumber\\&
-\left[\left(D_{\mu} \Phi^{\mu \nu}\right)^{*} D_{\nu} \Phi+\text { c.c. }\right] 
+2 R_{\mu \nu \rho \lambda} \Phi^{* \mu \rho} \Phi^{\nu \lambda}
\nonumber\\
& -\frac{1}{4} R|\Phi|^{2} -\text{i} q_{d} F_{\mu \nu} \Phi^{* \mu \lambda} \Phi_{\lambda}^{\nu}
\nonumber\\
& -m_{\mathrm{d}}^{2}\left(\left|\Phi_{\mu \nu}\right|^{2}-|\Phi|^{2}\right)-\lambda_{d}(\left|\Phi_{\mu \nu}\right|^{2}-|\Phi|^{2})^2~.
\end{align}
$\Psi$ is a complex scalar field and $\Phi_{\mu \nu}$ is the complex tensor fields with $\Phi=g^{\mu\nu}\Phi_{\mu\nu}$ denoting its trace. The covariant derivative act on the s-wave and d-wave fields as 
$D_{\mu}= \nabla_{\mu}-\text{i} q_{a} A_{\mu} \quad(a=\mathrm{s}, \mathrm{d})$, respectively. $\Lambda=-3/L^{2}$ is the negative cosmological constant, where L is the AdS radius. $\lambda_{s}$ and $\lambda_{d}$ are the coefficients of the fourth power self-interaction terms, which are believe to be powerful in tuning the potential landscape to obtain varies phase transitions~\cite{Zhang:2021vwp}. The interaction term with coefficient $\lambda_{sd}$ introduce the relevance between the s-wave and d-wave orders, and is also considered in Ref~\cite{Nishida:2014lta}. Working in the probe limit, we take the planar symmetric black brane solution~\cite{Lu:2020iav,Hennigar:2020lsl,Cai:2001dz} as the background spacetime
\begin{align}
	ds^{2}=\frac{L^{2}}{z^{2}}\left(-f(z) \text{d} t^{2}+\frac{\text{d} z^{2}}{f(z)}+\text{d} x^{2}+\text{d} y^{2}\right),
\end{align}
where the function f(z) is
\begin{align}
	f(z)=\frac{1}{2 \alpha z^{2}}\left[1-\sqrt{1-\frac{4 \alpha}{L}\left(1-\frac{z^{3}}{z_{h}^{3}}\right)}\right]~. \label{Functionf}
\end{align}
The relationship between this new Gauss-Bonnet coefficient $\alpha$ in (\ref{Functionf}) and the one in the action is given by $\alpha=\bar{\alpha}(D-3)(D-4)$. In the limit $D\rightarrow 4$, we need a finite $\alpha$ to get nontrivial solutions deformed by the higher curvature term, which leads an infinite $\bar{\alpha}$.
Near the AdS boundary $z\rightarrow 0$, the function f(z) gets the asymptotic value
\begin{align}
f(z)\rightarrow \frac{1}{2 \alpha z^{2}}\left(1-\sqrt{\frac{4 \alpha}{L^{2}}}\right),
\end{align}
therefore, the AdS radius is deformed to be
\begin{align}
L_{\text {eff }}^{2}=\frac{2 \alpha}{1-\sqrt{1-\frac{4 \alpha}{L^{2}}}} \rightarrow\left\{\begin{array}{ll}
L^{2}, & \text { for } \alpha \rightarrow 0 \\
\frac{L^{2}}{2}, & \text { for } \alpha \rightarrow \frac{L^{2}}{4}
\end{array}\right.~.
\end{align}
In order to maintain a real value for $L_{\mathrm{\mathrm{eff}}}^{2}$, we have $\alpha \leq L^{2} / 4$, which is called the Chern-Simons limit. From Eq.~(\ref{Functionf}) we see that $z=z_{h}$ is the location of the horizon. The temperature of the system corresponds to the Hawking temperature T of the black brane
\begin{align}
	T=\frac{3}{4 \pi z_{h}}.
\end{align}
We take the following ansatz
\begin{align}
\Psi=\psi(z), \quad \Phi_{x y}=\Phi_{y x}=\frac{L^{2}}{2 z^{2}} \varphi(z), \quad A_{t}=\phi(z),
\end{align}
with the other field components set to zero. Then the equations of motion are reduced to the following three
\begin{align}
\psi^{\prime \prime}+\left(\frac{f^{\prime}}{f}-\frac{2}{z}\right) \psi^{\prime}+\frac{q_{s}^{2} \phi^{2} \psi}{f^{2}}-\frac{m_{s}^{2} \psi}{z^{2} f}
&\nonumber\\
-\frac{2 \lambda_{s} \psi^{3}}{z^{2} f}-\frac{\lambda_{sd} \varphi^{2} \psi}{2 z^{2} f}&=0~, \\
\varphi^{\prime \prime}+\left(\frac{f^{\prime}}{f}-\frac{2}{z}\right) \varphi^{\prime}+\frac{q_{d}^{2} \phi^{2} \varphi}{f^{2}}-\frac{m_{d}^{2} \varphi}{z^{2} f}& \nonumber
\\
-\frac{\lambda_{d} \varphi^{3}}{z^{2} f}-\frac{\lambda_{sd}  \psi^{2}\varphi}{z^{2} f}&=0~,\label{a1.varphi}\\
\phi^{\prime \prime}-\frac{2 q_{s}^{2}\psi^{2} \phi}{z^{2} f}-\frac{q_{d}^{2} \varphi^{2} \phi}{z^{2} f}&=0~.
\end{align}
The Maxwell field $\phi$ should have a finite norm at the horizon, which implies $\phi|_{z=z_{h}}=0$. Therefore, the expansion of the three functions near the horizon are
\begin{align}
\psi(z)=&\psi_{0}+\psi_{1}\left(z-z_{h}\right)+\mathcal{O}\left(\left(z-z_{h}\right)^{2}\right), \\
\varphi(z)=&\varphi_{0}+\varphi_{1}\left(z-z_{h}\right)+\mathcal{O}\left(\left(z-z_{h}\right)^{2}\right),\\
\phi(z)=&\phi_{1}\left(z-z_{h}\right)+\mathcal{O}\left(\left(z-z_{h}\right)^{2}\right).
\end{align}
The charged scalar field $\psi(z)$ and the charged tensor field $\varphi(z)$ get natural boundary conditions at the horizon, therefore the derivative of the two functions are no longer independent and follows the constraints $\psi_{0}=-3 \psi_{1} /2$ and $\varphi_{1}=0$.

The expansions of these three fields near the AdS boundary are
\begin{align}
\psi=&\psi_s^- z+\psi_s^+ z^{2}~, \\
\varphi=&\varphi_d^-+\varphi_d^+ z^{3}~,\\
\phi=&\mu-\rho z~.
\end{align}
In this paper, we choose the standard quantization scheme, which means that $\psi_s^-$ and $\varphi_d^-$ are the sources of the s-wave and d-wave orders, while $\psi_s^+$ and $\varphi_d^+$ are the corresponding expectation values, respectively. $\mu$ is the chemical potential and $\rho$ is the charge density in the boundary field theory. We set the Dirichlet boundary conditions $\psi_s^-=\varphi_d^-=0$, which ensure the spontaneous breaking of the U(1) symmetry. Usually the charge parameters $q_s$ and $q_d$ do not play important role in the holographic models with single condensate because of the scaling symmetries in the equations of motion. However, the ratio $q_d/q_s$ give nontrivial contribution in the model with both the s-wave and d-wave orders. Therefore we set $q_s=1$ and leave the freedom of the value of $q_d$ to control the phase transitions. For convenience, we further set $m_d^2 L_{\mathrm{eff}}^2=0$, $m_s^2 L_{\mathrm{eff}}^2=-2$ and $L=z_h=1$.

In order to investigate the competition between the different solutions, it is necessary to compare the thermodynamic potential of these solutions. We work in the grand canonical potential in this work and calculate the grand potential from the on-shell Euclidean action. In the probe limit, the gravity background is fixed for the various solutions, and only the contribution from the matter part is important
\begin{align}
\Omega=&\frac{V_{2}}{T}\Big[-\frac{\mu \rho}{2}+\int \big(\frac{q_s^2\phi^2 \psi^2}{f z^2}+\frac{q_d^2\phi^2 \varphi^2}{2 z^2 f}
\nonumber\\
&-\frac{\lambda_s\psi^4}{z^4}-\frac{\lambda_d \varphi^4}{4 z^4}-\frac{\lambda_{sd} \psi^2 \varphi^2}{2 z^4}\big)\text{d}z\Big]~.
\end{align}
\section{The phase structure involving the s+d phase and the role of the fourth power nonlinear terms\label{sect:3}}
In order to investigate the phase structure involving the s+d phase in the Einstein-Gauss-Bonnet gravity, we need to first set the charge ratio $q_d/q_s$ to appropriate values to find out the s+d phase at nearly zero value of the Gauss-Bonnet coefficient. 
Then we fix $q_d/q_s$ to one appropriate value and draw the $\alpha-\mu$ phase diagram to present the influence of the Gauss-Bonnet term. After that, we also present the power of the fourth power nonlinear terms, and finally show the dependence of the special value of $\lambda_s$ and $\lambda_d$ on the value of the Gauss-Bonnet coefficient $\alpha$, which is important in realizing first order phase transitions in the single condensate s-wave and d-wave models.
\subsection{Locate the s+d phase with the appropriate charge ratio\label{subsection1}}
The value of the charge coupling $q_s$ and $q_d$ is usually set to 1 in holographic models with single condensate, because we can always apply the scaling symmetry to rescale the value of $q_s$ or $q_d$ to any finite value. However, in a holographic model with two orders or more, the scaling symmetry only shift all the charge couplings with the same proportion, while the ratios, such as $q_d/q_s$ in the s+d model, are stilled fixed. Therefore, without loss of generality, we set $q_s=1$ and take different values of $q_d$ to find out the s+d coexistent phase with $\alpha=1\times10^{-7}$ and zero values of the fourth power coefficients $\lambda_s$, $\lambda_d$ and $\lambda_{sd}$ on the first step.

\begin{figure}
\includegraphics[width=0.6\columnwidth]{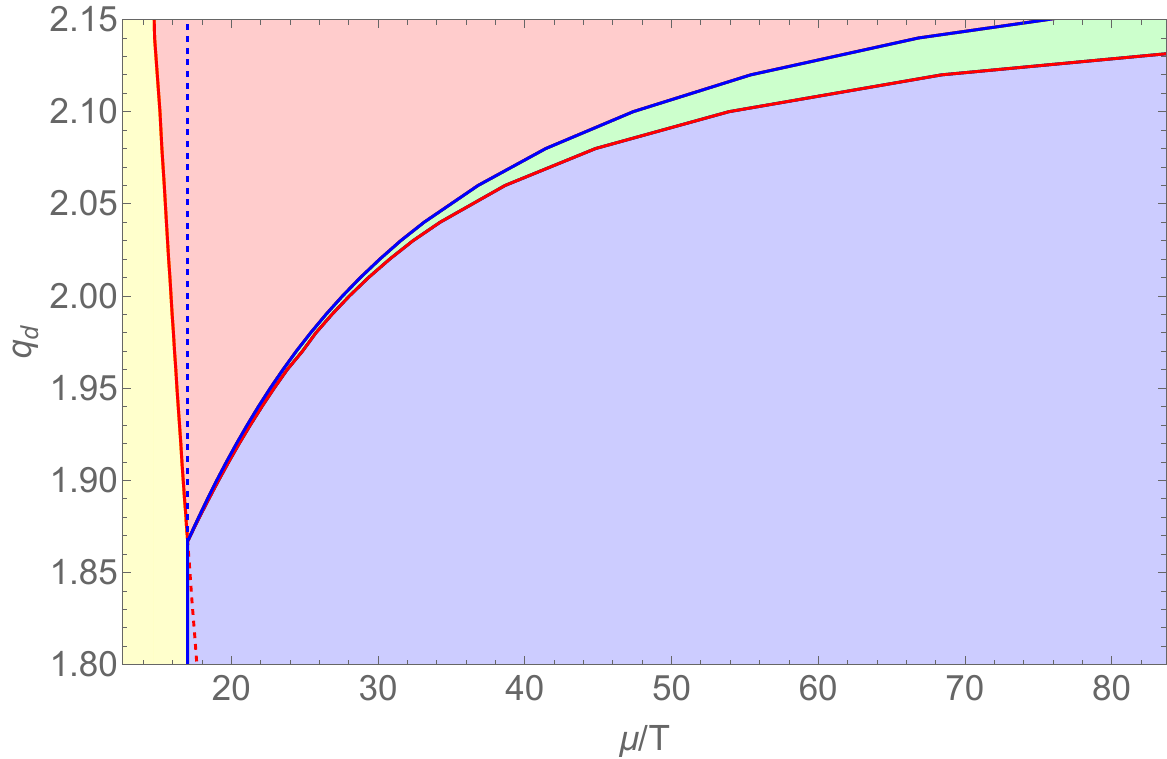}
\caption{The $q_{d}-\mu$ phase diagram with $\alpha=1\times10^{-7}$, $\lambda_{s}=\lambda_{d}=\lambda_{sd}=0$, $m_s^2 L_{\mathrm{eff}}^2=-2$, $m_d^2 L_{\mathrm{eff}}^2=0$. The yellow region is dominated by the normal phase, while the red, blue and green regions are dominated by the d-wave phase, the s-wave phase and the s+d coexistent phase, respectively. The red and blue lines represent the critical points of the d-wave and s-wave orders, respectively. 
\label{qdmuPD}}
\end{figure}

It should be noticed that, the s+d model is already studied in the Einstein gravity in Refs.~\cite{Nishida:2014lta,Li:2014wca}. However, the detailed phase structure with the same choice of the mass parameters $m_s^2 L_{\mathrm{eff}}^2=-2$ and $m_d^2 L_{\mathrm{eff}}^2=0$ is not presented. In Ref.~\cite{Nishida:2014lta}, only the special case with the charge ratio $q_d/q_s=1.95$ is investigated. In Ref.~\cite{Li:2014wca}, the detailed phase diagram of this s+d model is given with $m_s^2L^2=-2$ and $m_d^2L^2=7/4$. Therefore, in order to present our results more completely, we plot the $q_d-\mu$ phase diagram in Figure~\ref{qdmuPD}.

We can see from Figure~\ref{qdmuPD} that the s+d coexistent phase marked by the narrow green region only exist with $q_d>1.87$. Therefore in order to present the influences of the Gauss-Bonnet term with the coefficient $\alpha$ on the s+d phase, we choose $q_d=2.0$ in the next subsection.
\subsection{The phase diagram with a varying Gauss-Bonnet coefficient}\label{subsection2}
\begin{figure}
\includegraphics[width=0.6\columnwidth]{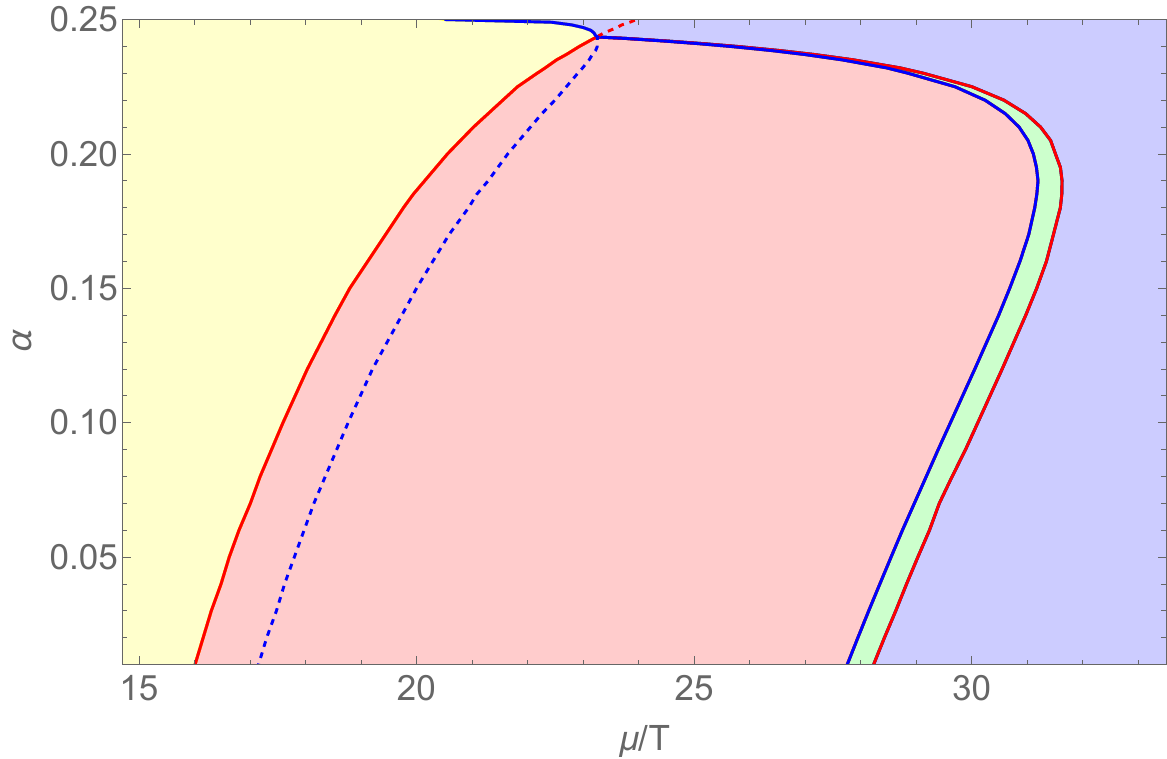}
\caption{The $\alpha-\mu$ phase diagram for the 4D Einstein-Gauss-Bonnet gravity. The parameters are set to the following values: $q_{d}=2.0$, $\lambda_{s}=\lambda_{d}=\lambda_{sd}=0$, $m_s^2 L_{\mathrm{eff}}^2=-2$, $m_d^2 L_{\mathrm{eff}}^2=0$. The colors in this plot are used as the same as in Figure~\ref{qdmuPD}.
} \label{GBmuPD}
\end{figure}

We known when $q_d>1.87$, the s+d phase exist between the s-wave phase and the d-wave phase. Here we fix $q_d=2.0$ and draw the $\alpha-\mu$ phase diagram in Figure~\ref{GBmuPD} to further show the influence of the Gauss-Bonnet term on the phase transitions involving the s+d phase.

In this phase diagram, we still use the red and blue lines to indicate the critical points of the d-wave order and the s-wave order respectively. The dashed section of the red and blue lines indicate the critical point of the unstable single condensate solution which get a higher value of the grand potential than the most stable phase. We can see from the two lines for the critical points of the single condensate solutions that along with the increasing of the Gauss-Bonnet coefficient, the critical chemical potential of the d-wave solution increases monotonically, while the critical chemical potential of the s-wave solution increases at small values of the Gauss-Bonnet coefficient, but decreases when the Gauss-Bonnet coefficient $\alpha$ approaches the Chern-Simons limit. This behavior of the s-wave phases with non zero values of the mass parameter $m_s^2 L_{\mathrm{eff}}^2$ is already observed in the studies on the s+s and s+p models in the Einstein-Gauss-Bonnet gravity~\cite{Li:2017wbi,Zhang:2023uuq}, and is explained by the change of $m_s^2$ when $m_s^2 L_{\mathrm{eff}}^2$ is fixed to have fixed value of the conformal dimension when the Gauss-Bonnet coefficient is tuned. It is also presented in Ref.~\cite{Qiao:2020hkx} that fixing the value of $m_s^2 L_{\mathrm{eff}}^2$ in the holographic studies from the Einstein-Gauss-Bonnet gravity is a better choice.

According to the sudden decreasing of the critical chemical potential near the Chern-Simons limit, the s-wave solution becomes more stable and the relation between the grand potential curves of the s-wave and p-wave solutions changes strongly near the intersection point of the red and blue lines. Therefore the region of the s+d phase also moves significantly to meet this intersection point, resulting in the big bend of the green region in the phase diagram. The above special properties in the $\alpha-\mu$ phase diagram indicates possible nontrivial $1/N$ effects when the field theory goes away from the large N limit.

\begin{figure}
\includegraphics[width=0.6\columnwidth]{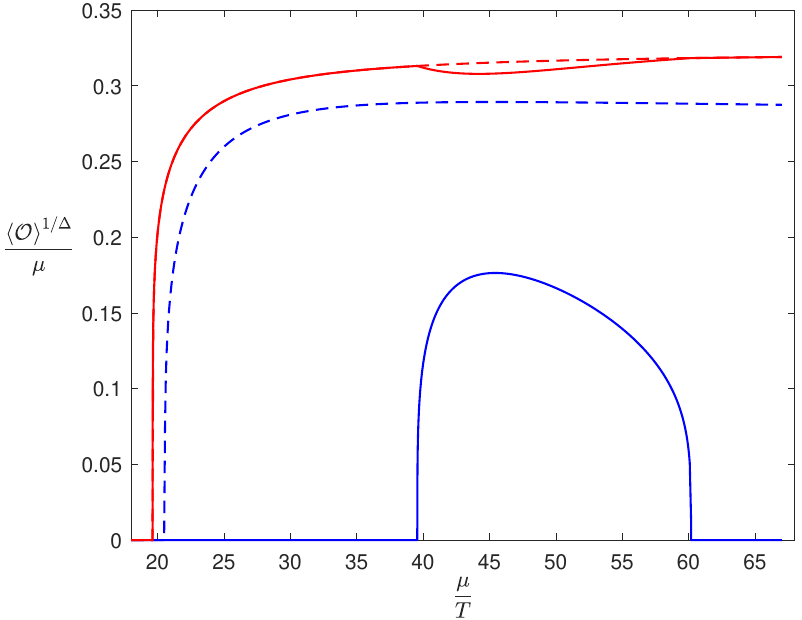}
\caption{The condensates in the case of the reentrance with $q_d=2.4455$, $\alpha=1/4$, $\lambda_{s}=\lambda_{d}=\lambda_{sd}=0$, $m_s^2 L_{\mathrm{eff}}^2=-2$, $m_d^2 L_{\mathrm{eff}}^2=0$. The red line denotes the condensate of the d-wave order, where the blue line denotes the condensate of the s-wave order. The dashed lines are for the unstable sections of superfluid solutions and the solid lines for the most stable solution.} \label{Reentrance}
\end{figure}

With this special behavior in the Einstein-Gauss-Bonnet gravity, the grand potential curve of the s-wave solution with $m_s^2 L_{\mathrm{eff}}^2=-2$ and that of the d-wave solution with $m_d^2 L_{\mathrm{eff}}^2=0$ changes very differently near the Chern-Simons limit. As a result, it is possible to find out non monotonic behavior of the condensates such as the reentrance. According to the systematic way described in Ref.~\cite{Li:2017wbi}, we also find a case showing the reentrant s-wave order with $q_d=2.4455$, $\alpha=1/4$ and we plot the condensates in Figure~\ref{Reentrance}.

From Figure~\ref{Reentrance} we can see that if we increase the chemical potential from a very small value, the system first undergoes a second order phase transition from the normal phase to the single condensate d-wave phase. After that, when the chemical potential goes on increasing, the system under goes the second phase transition from the single condensate d-wave phase to the s+d coexistent phase, and finally undergoes the third phase transition from the s+d phase and reenter the single condensate d-wave phase. This is a typical reentrent phase transition and is called the ``n-type'' in Ref.~\cite{Nie:2014qma} because of the shape of the s-wave condensate form a shape of the letter ``n''.
\subsection{The power of the fourth power terms and the special value dependence on the Gauss-Bonnet coefficient}\label{subsection3}
In this section, we investigate the power of the fourth power terms with coefficients $\lambda_{s}$, $\lambda_{d}$, and $\lambda_{sd}$.

\begin{figure}
\includegraphics[width=0.3\columnwidth]{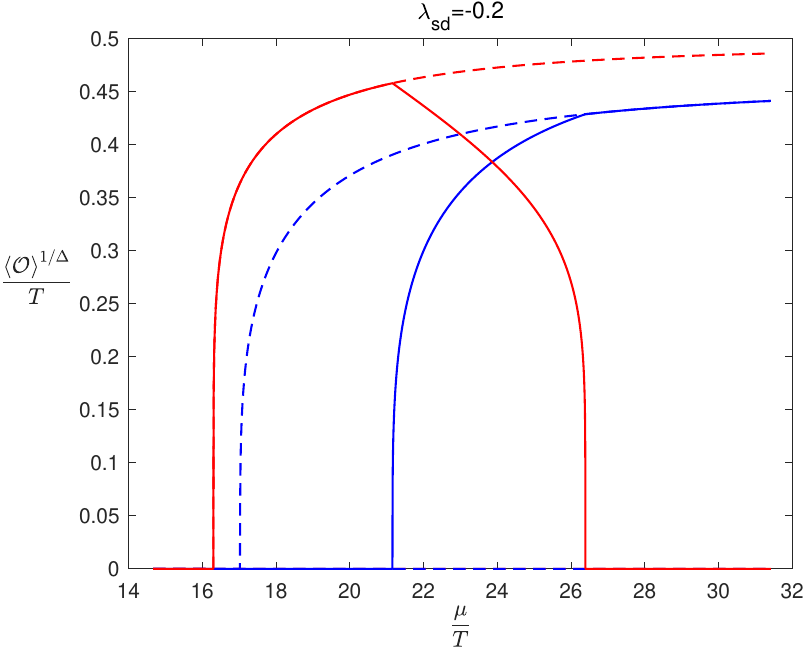}
\includegraphics[width=0.3\columnwidth]{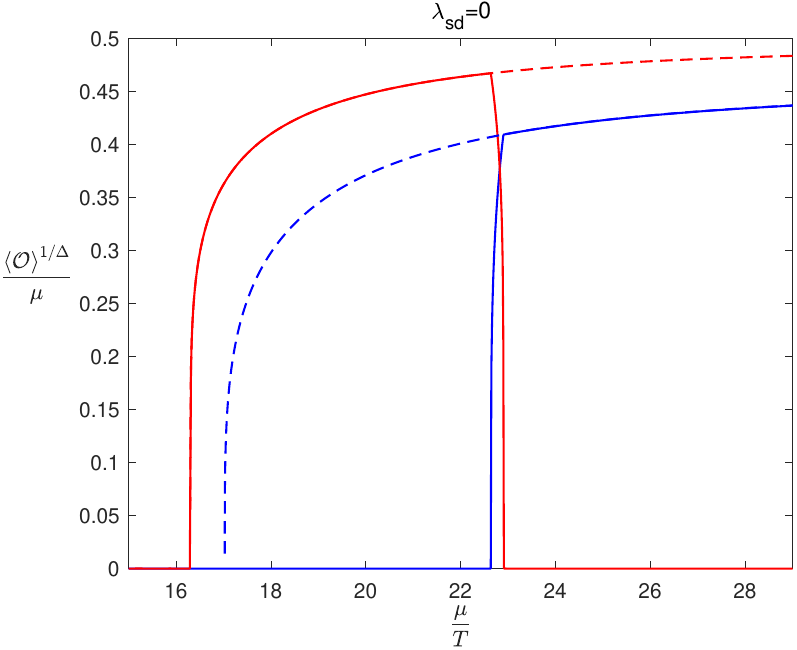}
\includegraphics[width=0.3\columnwidth]{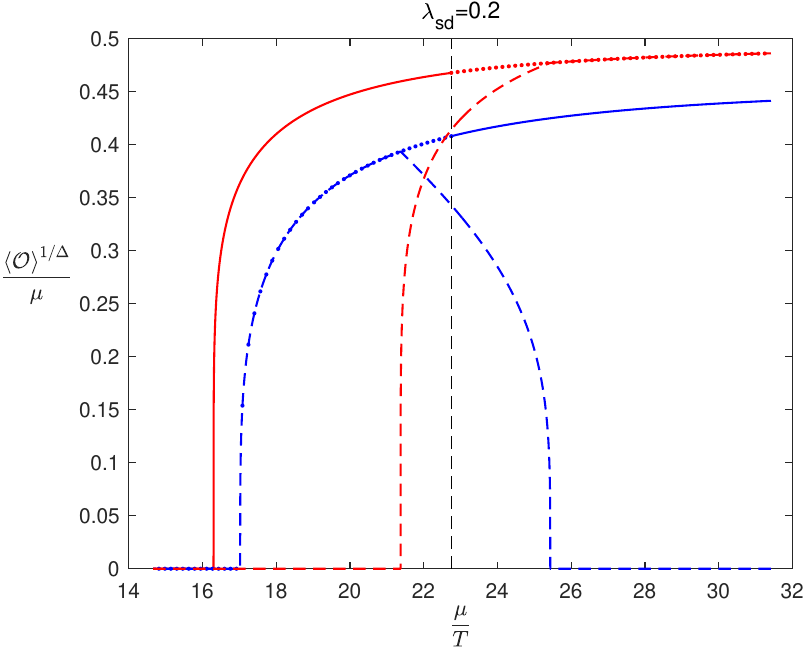}
\caption{The s-wave and d-wave condensates for $q_{d}=1.95$ and $\lambda_{sd}=-0.2$ (Left), $\lambda_{sd}=0$ (Middle), $\lambda_{sd}=0.2$ (Right), respectively. The red and blue curves represent the condensate of the d-wave order and the s-wave order, respectively. The solid sections are for the most stable phase while the dotted sections are for unstable solutions. This figure represents the power of $\lambda_{sd}$ initially studied in Ref.~\cite{Nishida:2014lta} in a more complete manner.} \label{CondensateLambdaSD}
\end{figure}

From the coupled terms in the equations of motion, we can see that varying $\lambda_{sd}$ would not change the single condensate s-wave solutions and d-wave solutions. Only the s+d coexistent solutions depend on the value of $\lambda_{sd}$. The detailed influences of $\lambda_{sd}$ is already studied in Ref.~\cite{Nishida:2014lta} in Einstein gravity, and is represented here with more complete condensate curves for a very small Gauss-Bonnet coefficient $\alpha=1\times 10^{-7}$ and $q_d=1.95$ in Figure.~\ref{CondensateLambdaSD}, where the three panels showing condensate curves with $\lambda_{sd}=-0.2$ (Left panel), $0$ (Middle panel) and $0.2$ (Right panel), respectively. From these plots we confirm that the value of $\lambda_{sd}$ do not change the condensates of the single condensate solutions at all. With a lower value of $\lambda_{sd}$ the width of the coexistent s+d phase is larger, while with a higher value of $\lambda_{sd}$, the width of the stable s+d phase becomes smaller. It seems that the width of the s+d solution in the right panel with $\lambda=0.2$ is larger than the s+d phase in the middle panel with $\lambda=0$, but notice that in the case of $\lambda=0.2$, the s+d solution becomes totally unstable and a first order phase transition is expected between the s-wave phase and the d-wave phase at the phase transition point marked by the vertical dotted black line.

\begin{figure}
\includegraphics[width=0.6\columnwidth]{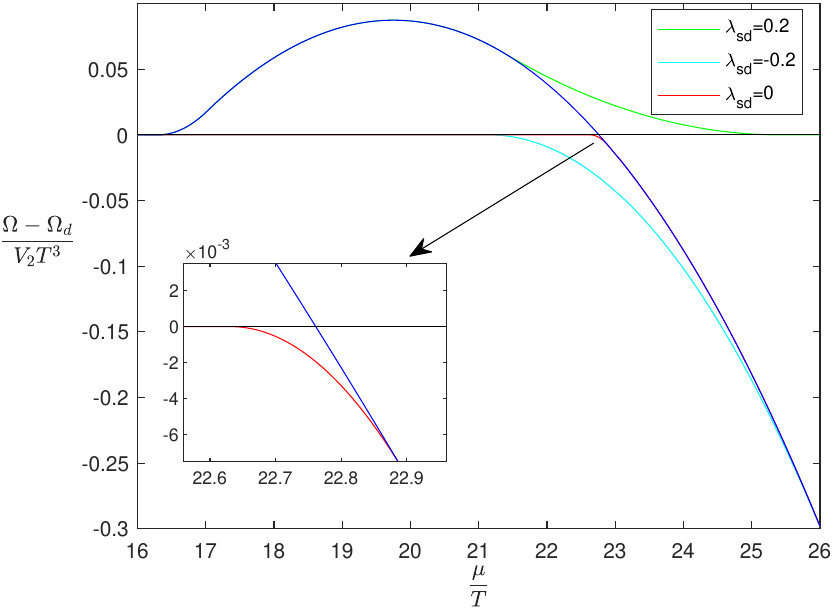}
\caption{The grand potential curves for the s+d solutions with $q_{d}=1.95$ and $\lambda_{sd}=-0.2$ (cyan), $\lambda_{sd}=0$ (red), $\lambda_{sd}=0.2$ (green), respectively. The bottom left corner shows a zoomed-in view for the s+d solution with $\lambda_{sd}=0$ (red). In this figure, we plot relative value with respect to the single condensate d-wave solutions, which are indicated by the horizontal black line. The solid blue line indicates the single condensate s-wave solutions.} \label{OmegaLambdaSD}
\end{figure}

It is also necessary to plot the grand potential lines to confirm the stability relations, and we choose to plot the relative values with respect to the d-wave solution in Figure~\ref{OmegaLambdaSD}, where the grand potential line of the d-wave solution is along the horizontal axes. The solid blue line indicating the grand potential curve for the s-wave solution do not change with the different values of $\lambda_{sd}$ and thus the curves of the s+d solutions for three different values of $\lambda_{sd}$ are able to be presented in the same panel very clearly. The cyan line for the s+d solution with $\lambda_d=-0.2$ is lower than the red line for the s+d solution with $\lambda_d=0$. The green line for the s+d solution with $\lambda_d=0.2$ form a standard swallow tail shape with the blue and black curves, indicating the instability of these s+d solutions on the green line.

It is obvious from the equations of motion that the different values of $\lambda_d$ do not change the condensate as well as the grand potential of the single condensate s-wave solutions. In the same way, the different values of $\lambda_s$ do not change the condensate as well as the grand potential of the single condensate d-wave solutions, either. The influence of $\lambda_{s}$ on the single condensate s-wave solutions is well studied in Refs.~\cite{Zhang:2021vwp,Zhao:2022jvs} in Einstein gravity. We expect that the influence of $\lambda_d$ on the single condensate d-wave solutions should be similar to the s-wave cousin. Therefore, we focus on studying the influence of $\lambda_d$ on the single condensate d-wave solutions which still has not been investigated previously, and confirming the universality.

Because the theoretical analyze on the equations of motion~\cite{Zhang:2021vwp} as well as the thermodynamic potential landscape~\cite{Zhao:2022jvs} are quite universal, we fix the Gauss-Bonnet coefficient $\alpha=1\times 10^{-7}$ to present the influence of $\lambda_d$ on the single condensate d-wave solutions without loss of generality. We choose such a small value for the Gauss-Bonnet coefficient, that the quantitative results are also useful to help understand holographic models with a d-wave order in the Einstein gravity.

Near the critical points of the d-wave order, the d-wave field $\varphi$ is infinitesimal, ensuring the equation of $\varphi$ to be linearized and the nonlinear terms ignored, therefore the critical point of the single condensate d-wave phase would not be affected. Even with considering the s+d phase, the critical point of the d-wave order for the s+d phase is not altered, because firstly the background s-wave phase is not depend on $\lambda_d$, and secondly near the critical point of the d-wave order, the nonlinear terms are still ignorable in the equations of $\varphi$. Similar laws are confirmed in the holographic s+p model with fourth power terms by the numerical results~\cite{Zhang:2021vwp}.
\begin{figure}
\includegraphics[width=0.4\columnwidth]{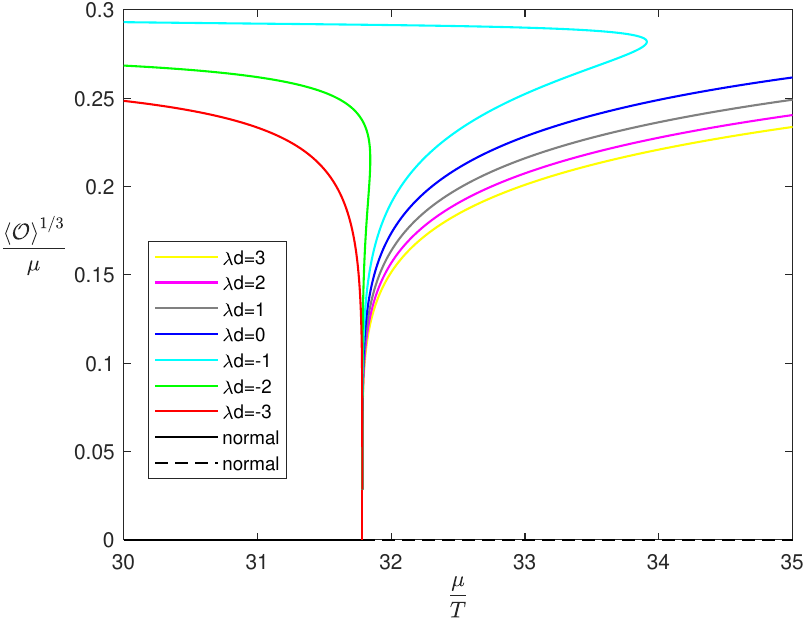}
\includegraphics[width=0.4\columnwidth]{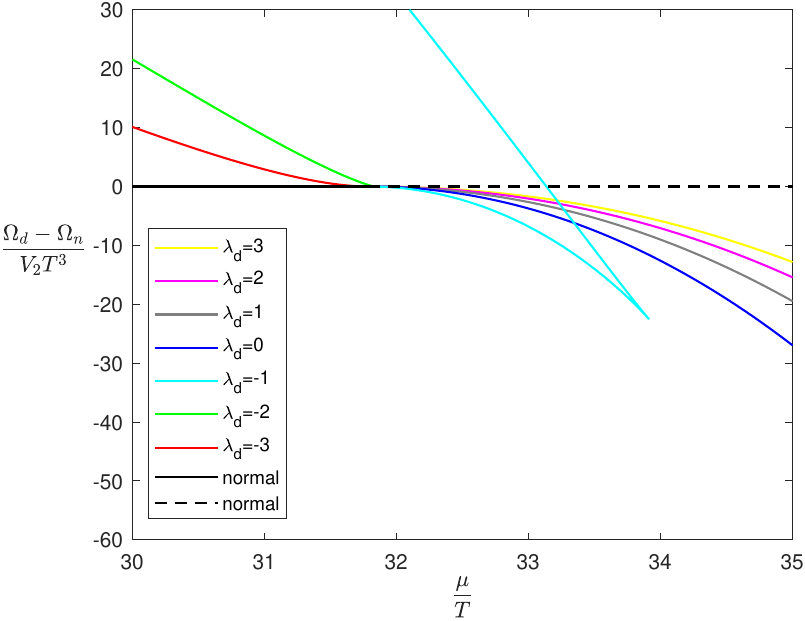}
\caption{The condensates (Left) and the grand potential curves (Right) for the single condensate d-wave solutions with various values of $\lambda_d$. We set $\alpha=1\times 10^{-7}$, $q_d=1$ and $\lambda_{s}=\lambda_{sd}=0$. Colored lines are used to present the various values of $\lambda_d$ as $\lambda_{d} =3$ (yellow), $\lambda_{d} =2$ (purple), $\lambda_{d}=1$ (gray), $\lambda_{d}=0$ (blue), $\lambda_{d}=-1$ (cyan), $\lambda_{s}=-2$ (green), $\lambda_{s}=-3$ (red).} \label{lambdaD}
\end{figure}

We show the condensates as well as the grand potential curves for the single condensate d-wave solutions with $\alpha=1\times 10^{-7}$, $q_s=q_d=1$, $\lambda_{s}=\lambda_{sd}=0$ and various values of $\lambda_d$ in the left and right panels of Figure~\ref{lambdaD}. We can see from these two plots that compared to the d-wave solution with $\lambda_d=0$ colored by the blue line, a increasing positive value of $\lambda_d$ will reduce the condensate and increase the grand potential. However, when $\lambda_d$ is lowered to a slightly negative value, the change is not only quantitatively make the condensate larger and the grand potential lower, but also make the condensate curve growing to a different direction at a large value of condensate, resulting in a turning point in the condensate curve as well as the grand potential line, which is a signal of the 0th order phase transition~\cite{Cai:2013aca,Zhao:2022jvs}. When we further lower the value of $\lambda_d$, the condensate curve grows leftward even at the critical point, as indicated by the red curve in the left panel of Figure~\ref{lambdaD}. From the red curve in the right panel, the d-wave phase in this case is totally unstable. The similar behavior of the single condensate s-wave solution is already observed in Ref.~\cite{Zhao:2022jvs}, and the special value of $\lambda_s$, below which the condensate grows to an opposite direction at the critical point, is important in models with additional 6th power terms, when one needs to realize a 1st order phase transition~\cite{Zhao:2024jhs}. The special value of $\lambda_s$ was confirmed to be equal to $\Tilde{\lambda}_s=-1.949$ in the canonical ensemble~\cite{Zhao:2022jvs}, and similar special value exist for $\lambda_d$.

\begin{figure}
\includegraphics[width=0.6\columnwidth]{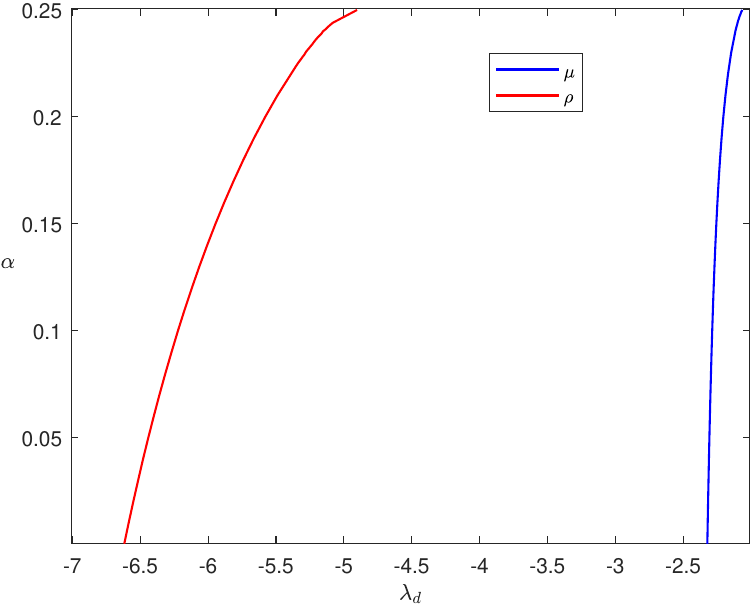}
\caption{The dependence of the special values $\Tilde{\lambda}_{d}$ on the Gauss-Bonnet coefficient $\alpha$. The solid blue line is for the case in the canonical ensemble where the charge density $\rho$ is fixed, while the solid red line is for the case in the grand canonical ensemble where the chemical potential $\mu$ is fixed. We set $q_{d}=1$, $m_s^2 L_{\mathrm{eff}}^2=-2$ and $m_d^2 L_{\mathrm{eff}}^2=0$ in this plot.} \label{lambdaSpecial}
\end{figure}

We calculate the special value of $\lambda_d$ in both the canonical ensemble and the grand canonical ensemble, and plot their dependence on the Gauss-Bonnet coefficient $\alpha$ in Figure~\ref{lambdaSpecial}, where the solid red line is for the grand canonical ensemble and the solid blue line is for the canonical ensemble. We can see that the special value $\Tilde{\lambda}_d^\mu$ in the grand canonical ensemble is always lower than the special value $\Tilde{\lambda}_d^\rho$ in the canonical ensemble. Between the two special values, the section of the superfluid solutions with lower free energy in the canonical ensemble will suffer instabilities of inhomogeneous perturbations~\cite{Zhao:2022jvs,Zhao:2023ffs}.
\section{summary and prospect}\label{sect:4}
In this paper, we studied the phase transitions of a holographic s+d model with fourth power terms in the Einstein-Gauss-Bonnet gravity. We first set a very small value for the Gauss-Bonnet coefficient $\alpha=1\times10^{-7}$ to present the $q_d-\mu$ phase diagram, which is almost the same to the phase diagram in the model from Einstein gravity. The we choose an appropriate value $q_p=2.0$ to show the influence of the Gauss-Bonnet term on the phase transitions with a $\alpha-\mu$ phase diagram. Since the Gauss-Bonnet term is expected to bring in the large N corrections on the CFTs, our results show the possible influence of the $1/N$ effects on the phase diagram with coexisting s+d phase. We also present the power of the fourth power interaction terms. The influence of the different values of $\lambda_d$ on the phase transitions of the single condensate d-wave solutions follows the universal laws similar to the influence of $\lambda_s$ on the single condensate s-wave solutions. We finally present the dependence of the special values of $\lambda_d$, below which the phase transitions of the single condensate d-wave solutions grows to a different direction at the critical point, on the Gauss-Bonnet coefficient in both the canonical ensemble and the grand canonical ensemble. The two special values are important to realize 1st order phase transitions as well as study the inhomogeneous instabilities triggering the spinodal decompositions. The various superfluid phase transitions including the 1st order and the re-entrant ones also indicate that more complex phase transitions involving the s-wave and d-wave orders are possible and should be more carefully explored in real cuprate superconductors.

There are many possible interesting future studies based on this work. For example, it is interesting to include the 6th power terms for the d-wave order to study the inhomogeneous instabilities as well as the non equilibrium processes in 1st order d-wave phase transitions. It is also necessary to include the back-reaction of the matter fields on the metric to study interesting features, such as the entanglement structure, the inner geometry as well as the hydrodynamics for the holographic s+d models. Linear perturbations in the s+d model in the probe limit will also present more rich physics because of different symmetries of the superfluid ground states involving the s-wave and d-wave orders.
\section*{Acknowledgements}
This work is partially supported by the National Natural Science Foundation of China (Grant Nos.11965013).
ZYN is partially supported by Yunnan High-level Talent Training Support Plan Young $\&$ Elite Talents Project (Grant No. YNWR-QNBJ-2018-181).
\bibliographystyle{unsrt}
\bibliography{reference}
\end{document}